\documentclass[twocolumn,prb,amsmath,amssymb,floatfix,superscriptaddress,showpacs]{revtex4}
\usepackage{graphicx}
\usepackage{dcolumn}
\usepackage{bm}
\usepackage{times}
\begin{document}

\title{ Effect of magnetic field on the spin resonance in FeTe$_{0.5}$Se$_{0.5}$ as seen via inelastic neutron scattering}
\author{Jinsheng Wen}
\affiliation{Condensed Matter Physics and Materials Science
Department, Brookhaven National Laboratory, Upton, New York 11973,
USA} \affiliation{Department of Materials Science and Engineering,
Stony Brook University, Stony Brook, New York 11794, USA}
\author{Guangyong Xu}
\affiliation{Condensed Matter Physics and Materials Science
Department, Brookhaven National Laboratory, Upton, New York 11973,
USA}
\author{Zhijun Xu}
\affiliation{Condensed Matter Physics and Materials Science
Department, Brookhaven National Laboratory, Upton, New York 11973,
USA} \affiliation{Physics Department, The City College of New York,
New York City, New York 10031, USA}
\author{Zhi Wei Lin}
\affiliation{Condensed Matter Physics and Materials Science
Department, Brookhaven National Laboratory, Upton, New York 11973,
USA}
\author{Qiang Li}
\affiliation{Condensed Matter Physics and Materials Science
Department, Brookhaven National Laboratory, Upton, New York 11973,
USA}
\author{Ying Chen}
\affiliation{NIST Center for Neutron Research, National Institute of
Standards and Technology, Gaithersburg, Maryland 20899, USA}
\author{Songxue Chi}
\affiliation{NIST Center for Neutron Research, National Institute of
Standards and Technology, Gaithersburg, Maryland 20899, USA}
\author{Genda Gu}
\affiliation{Condensed Matter Physics and Materials Science
Department, Brookhaven National Laboratory, Upton, New York 11973,
USA}
\author{J.~M.~Tranquada}
\affiliation{Condensed Matter Physics and Materials Science
Department, Brookhaven National Laboratory, Upton, New York 11973,
USA}
\date{\today}

\begin{abstract}
Inelastic neutron scattering and susceptibility measurements have
been performed on the optimally-doped Fe-based superconductor
FeTe$_{0.5}$Se$_{0.5}$, which has a critical temperature, $T_c$ of
14~K. The magnetic scattering at the stripe antiferromagnetic
wave-vector ${\bf Q}=(0.5,0.5)$ exhibits a ``resonance'' at
$\sim6$~meV, where the scattering intensity increases abruptly when
cooled below $T_c$. In a 7-T magnetic field parallel to the $a$-$b$
plane, $T_c$ is slightly reduced to $\sim12$~K, based on
susceptibility measurements. The resonance in the neutron scattering
measurements is also affected by the field. The resonance intensity
under field cooling starts to rise at a lower temperature
$\sim12$~K, and the low temperature intensity is also reduced from
the zero-field value. Our results provide clear evidence for the
intimate relationship between superconductivity and the resonance
measured in magnetic excitations of Fe-based superconductors.

\end{abstract}


\maketitle

The recent discovery of of Fe-based
superconductors~\cite{hosono_2,hosono4,chen-2008-453,ren-2008-25,hsu-2008,
yeh-2008} has triggered tremendous interest in the field. One of the
key questions to be answered is what is the pairing mechanism for
the high critical temperature (high-$T_c$) superconductivity in
these materials. It is now widely believed that pairing mediated by
magnetic excitations is the most likely candidate for explaining the
superconductivity.~\cite{mazin:057003,kuroki-2008,ma:033111,dong-2008-83,cvetkovic-2009,graser-2009}
The ``resonance'' in magnetic excitations, where the spectral weight
at the resonance energy shows a significant increase when the system
enters the superconducting phase, has been observed in a number of
these Fe-based superconductors, including BaFe$_2$As$_2$ (the 1:2:2
system)~\cite{christianson-2008-456,lumsden:107005,chi:107006,shiliang-2009,
inosov-2009} and the 1:1 system
Fe$_{1+\delta}$Te$_{1-x}$Se$_x$~\cite{qiu:067008, mook-2009}. The
resonance is always observed at the energy $\hbar\Omega_0 \sim
5k_BT_c$, and near the antiferromagnetic $(0.5,0.5)$ point (using
notation with two Fe atoms per unit cell) although the propagating
vectors for the spin-density-wave (SDW) in the parent compounds are
different by 45$^\circ$ in these two
systems.~\cite{bao-2009,li-2009-79,wen:104506} These results suggest
that the resonance in the magnetic excitations should be similar
across different Fe-based superconductor systems, and are closely
related to the onset of superconductivity.

In these superconductors, angle resolved photoemission (ARPES)
studies~\cite{liu:177005,arpesnature,fermisurfacepnas} have provided
evidence for electron and hole pockets that are nearly nested by the
stripe antiferromagnetic
wave-vector.~\cite{mazin:057003,seo:206404,mazin-2009}  A spin
resonance detectable by neutron scattering is predicted to occur at
a particular wave-vector only if that wave-vector connects portions
of the Fermi surface that have opposite signs of the superconducting
gap, so that observations of the resonance may provide important
information relevant to the symmetry of the superconducting
gap.~\cite{maier:020514,maier:134520} Since superconductivity, and
hence the pairing, is sensitive to magnetic field,  one would
naturally expect that an external magnetic field can also impact the
resonance accordingly, as seen in
YBa$_2$Cu$_3$O$_{6.6}$~(Ref.~\onlinecite{dairesonancefield}) and in
La$_{1.82}$Sr$_{0.18}$CuO$_4$ (Ref.~\onlinecite{tranquada-2004}).
Indeed, the magnetic field effect on the resonance in Fe-based
superconductors has been observed in the 1:2:2 system
BaFe$_{1.9}$Ni$_{0.1}$As$_2$,\cite{zhao-2009} where the resonance
energy and intensity have been partially reduced by an external
field.

We have carried out an inelastic neutron scattering study on an
optimally-doped 1:1 material---a single crystal of
FeTe$_{0.5}$Se$_{0.5}$, with $T_c\approx$14~K. We find that a
resonance with energy $\hbar\Omega_0\approx6$~meV$=5k_BT_c$ appears
below $T_c$, consistent with previous
findings.~\cite{qiu:067008,mook-2009} In a 7-T magnetic field
parallel to the $a$-$b$ plane, the superconductivity is partially
suppressed, with reduced $T_c$ of 12~K. In the field, the resonance
starts to appear at the reduced $T_c$, with lower intensity than
that measured in zero field. This behavior demonstrates that the
magnetic excitations have a close association with the
superconductivity.

\begin{figure}[ht]
\includegraphics[width=0.8\linewidth]{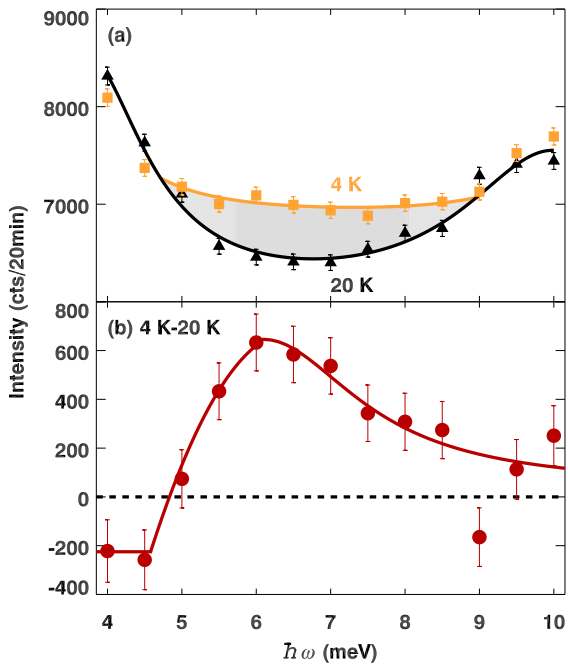}
\caption{(Color online) (a) Constant {\bf Q} scans at $(0.5,0.5,0)$
for temperatures below ($T=4$~K) and above ($T=20$~K) $T_c$. Shading
indicates the difference between scans.  (b) Data obtained by
subtracting 20~K data from 4~K data. Error bars represent square
root of total counts. Lines through data are guides for the eye. }
\label{fig:1}
\end{figure}

\begin{figure}[ht]
\includegraphics[width=0.8\linewidth]{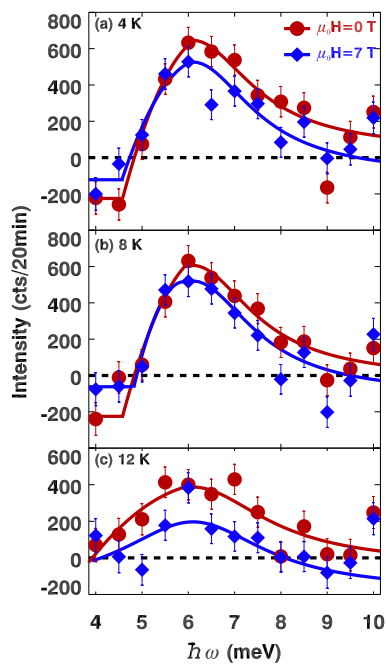}
\caption{(Color online) Constant {\bf Q} scans at (0.5,0.5,0), after
subtraction of the zero-field scan at 20~K. (a) $T=4$~K, (b) 8~K,
(c) 12~K, for $\mu_0 H= 0$~T (circles), and 7~T (diamonds). Error
bars represent square root of total counts. Lines through data are
guides for the eye.} \label{fig:2}
\end{figure}

\begin{figure}[ht]
\includegraphics[width=0.8\linewidth]{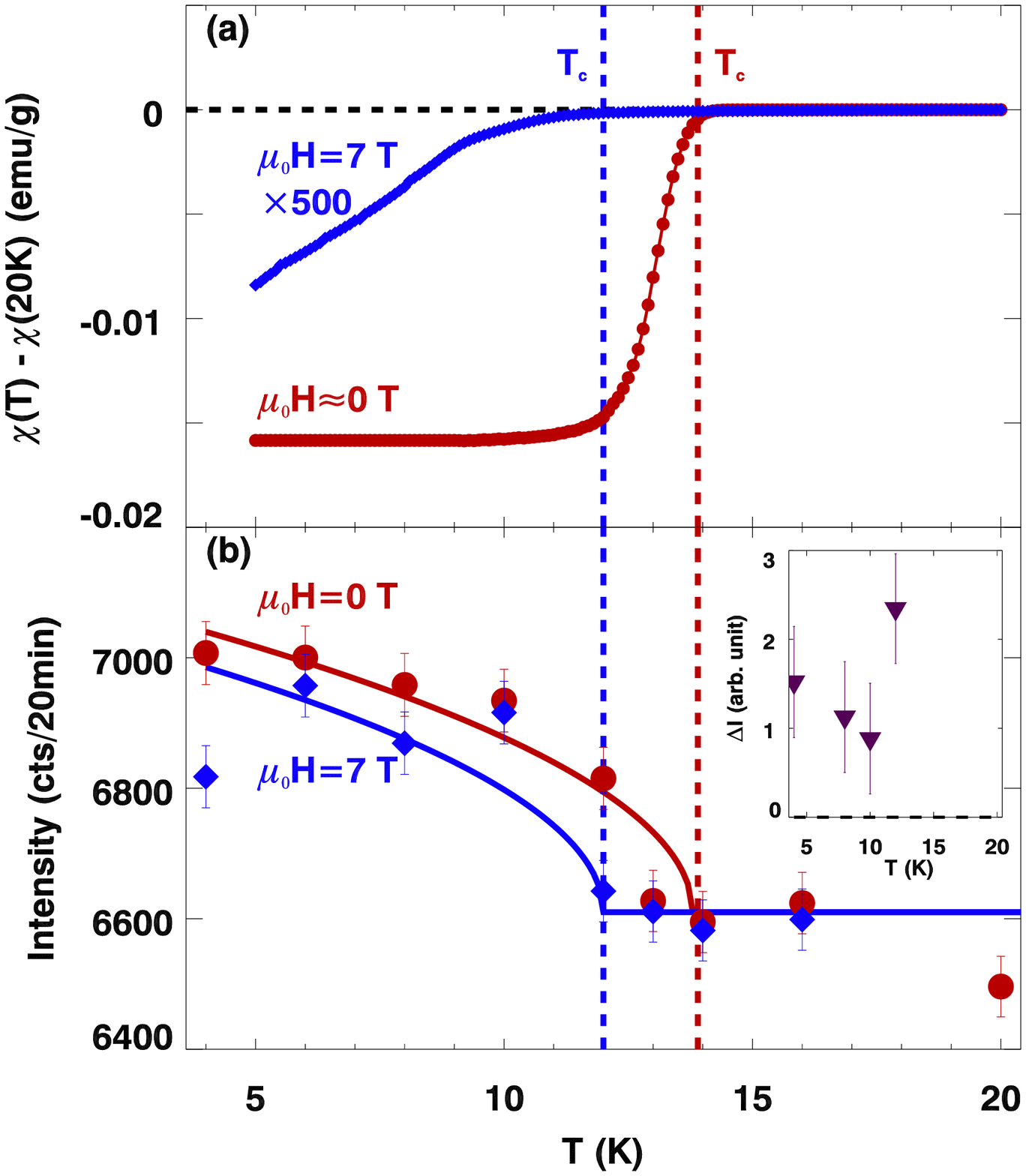}
\caption{(Color online) (a) Susceptibility measured with $\mu_0 H
=0.0005$~T (circles) and 7~T (diamonds), with field parallel to the
$a$-$b$ plane. Dashed lines indicate the $T_c$s. (b) Resonance
intensity at $(0.5,0.5,0)$ integrated from 6~meV to 7~meV. The solid
lines are fits using mean-field theory (described in the text), with
$T_c$s obtained from (a). Inset shows the difference of the
resonance intensities for 0~T and 7~T, integrated from 5~meV to
8~meV.  Error bars represent square root of total counts.}
\label{fig:3}
\end{figure}

The single-crystal sample was grown by a unidirectional
solidification method with nominal composition of
FeTe$_{0.5}$Se$_{0.5}$. The bulk susceptibility was characterized
using a superconducting quantum interference device (SQUID)
magnetometer. In the susceptibility measurements, the sample was
oriented so that $a$-$b$ plane was parallel to the magnetic field.
Neutron scattering experiments were carried out on the triple-axis
spectrometer BT-7 located at the NIST Center for Neutron Research. A
single crystal with mass of 8.9~g was used in the neutron experiment
and firmly fixed to an aluminum plate.  The lattice constants are
$a=b=3.80(8)$~\AA, and $c=6.14(7)$~\AA\ using the notation where
there are two Fe atoms in one unit cell. The data were collected in
$(HHL)$ scattering plane, defined by two vectors [110] and [001],
and described in reciprocal lattice units (r.l.u.) of $(a^*, b^*,
c^*)=(2\pi/a,2\pi/b,2\pi/c)$. A vertical magnetic field of 7~T was
applied parallel to the $a$-$b$ plane (along [1$\bar1$0]) in the
field-cooling (FC) measurements.

Energy scans have been performed at ${\bf Q}=(0.5,0.5,0)$, as shown
in Fig.~\ref{fig:1}(a). There is a large background at low energies
coming from the superconducting magnet in which the sample resides,
and this obscures the magnetic response in the raw data. However, if
we compare the scans taken at 4~K and 20~K, a significant amount of
spectral weight shows up between 5~meV and 9~meV for the spectrum
measured at low temperature (as indicated by the shading). If we
subtract the 20~K data from the 4~K data as in Fig.~\ref{fig:1}(b),
one can see a broad peak at $\sim6$~meV. This is consistent with
that observed in 40\% and 50\% Se doped samples, in which resonance
energies of 6.5~meV and 7~meV, respectively, were
reported.~\cite{qiu:067008,mook-2009}  Although a spin gap is not
directly observed in the raw data, we do see from the background
subtracted data in Fig.~\ref{fig:1}(b) that the difference of the
intensity ($I_{4K}-I_{20K}$) becomes negative below 5~meV, which
suggests that a gap opens below this energy at 4~K, consistent with
the gap value obtained by Qiu {\it et al.}.~\cite{qiu:067008}

To test the impact of a magnetic field, a 7-T field was applied at
20~K, and the sample was cooled in the field.  In Fig.~\ref{fig:2},
we show background (20~K data, zero field) subtracted scans
performed at different temperatures. At $T=12$~K, the difference
between data taken with and without the field is very clear. With
further cooling, the difference is still observable but becomes less
pronounced. At $T=4$~K, the peak intensity for the 7-T scan is about
10\% to 20\% smaller than that of the zero-field data, while the 7-T
spectrum seems to have more intensity filled in below the gap ($\sim
5$~meV).

We also performed some constant-energy (7-meV) scans along $(h,h,0)$
through $h=0.5$. With a counting rate of 5 min/point, the change in
signal at $h\approx0.5$ between 4~K and 20~K was consistent with the
constant-{\bf Q} scans; however, the signal-to-background level at
this counting rate was not sufficient to a provide a useful measure
of the peak shape, nor to resolve changes due to field. Given finite
beam time, it was not possible to measure both constant-{\bf Q} and
constant-energy scans with adequate statistics, so we decided to
abandon the latter.

There is a sum rule for scattering from spin-spin correlations, and
hence one might expect that the reduction of the resonance intensity
by the field should result in an increase of spectral weight below
the gap, as commonly seen in
cuprates,~\cite{science1759,kofu-2009-102,wen:212506,PLCCO_resonancefield}
as well as in
BaFe$_{1.9}$Ni$_{0.1}$As$_2$~(Ref.~\onlinecite{zhao-2009}). As
discussed above, it is consistent with our results in principle, but
the large background makes it impossible to follow the behavior to
lower energies. In cuprates, Demler {\it et al.} analyzed a model of
coexisting but competing phases of superconductivity and SDW
order,~\cite{prl67202} and successfully predicted the field-induced
static magnetic order observed
experimentally.~\cite{nature299,haug:017001,lakefield} We have
searched for SDW order around (0.5,0.5,0), but no evidence of such
field-induced order was found.

We have measured the bulk susceptibility in 0-T and 7-T field as
well, and the results are shown in Fig.~\ref{fig:3}(a). In zero
field, the system enters a superconducting state at 14~K, and
becomes fully diamagnetic below 12~K. In the 7-T field,
superconductivity is partially suppressed, and $T_c$ has been
reduced to 12~K. As a result of the suppressed superconductivity,
the resonance intensity has also been reduced as shown in
Fig.~\ref{fig:2}.

Fig.~\ref{fig:3}(b) gives another perspective of the impact of the
field on the resonance. There we plot the intensity,  integrated
from 6~meV to 7~meV, as a function temperature obtained for the
measurements with and without the field. The intensity $I(T)$ was
fit with the mean-field theory~\cite{solidstatephys} using $T_c$s
determined by the onset of the diamagnetism in Fig.~\ref{fig:3}(a),
with $I(T)=I(0)(1-T/T_c)^{1/2}+A$, where $I(0)$, and $A$ are
constants. This formula results in the solid lines, which fit the
data reasonably well. In both 0~T and 7~T, the resonance intensity
starts to appear below respective $T_c$, and increases with cooling.
At low temperatures, the intensity at 7 T is lower than the
zero-field value. To confirm that the intensity is reduced at 7 T,
we plot in the inset of Fig.~\ref{fig:3}(b) the difference between
intensity at 0~T and 7~T, $\Delta I$, integrated from 5~meV to
8~meV;  one can see that the intensity difference is well above
zero.

With Fig.~\ref{fig:3}, one can better understand the results in
Fig.~\ref{fig:2}, especially the most pronounced field effect at
12~K. In zero field, the sample is in superconducting state at 12~K,
where the resonance has finite intensity; in the 7-T field, the
system is driven to normal state at this temperature, and the
resonance intensity is approaching background level.

From the data, it is clear that the magnetic field depresses the
superconductivity, and also reduces the onset temperature and
intensity of the resonance. In principle, if the resonance is
directly associated with the superconducting volume of the sample,
the intensity ratio $I_{7T}/I_{0T}$ should be roughly proportional
to $1-H/H_{c2}$, where $H$ is the applied field, and $H_{c2}$ is the
upper critical field.~\cite{dairesonancefield} Our results showing a
change $\sim10\%$ in the resonance intensity, suggesting that
$H_{c2}$ is $\sim70$~T, which is comparable to the range estimated
in other studies.~\cite{si:052504,kida-2009} Although no significant
change in the resonance with field was identified for the 40\%\ Se
sample in Ref.~\onlinecite{qiu:067008}, we believe that our results
are consistent with that study within the error bars. The fact that
the field also suppresses the resonance intensity in
BaFe$_{1.9}$Ni$_{0.1}$As$_2$~\cite{zhao-2009} suggests that this
should be common in Fe-based superconductors.

There are of course, still issues not fully resolved based on our
results. For example, the quality of our data does not allow us to
accurately determine the resonance energy. It is therefore hard to
find out whether the resonance energy can be affected by the
external magnetic field or not, although it has been shown that the
former is the case in BaFe$_{1.9}$Ni$_{0.1}$As$_2$.~\cite{zhao-2009}
We have measured the susceptibility with field perpendicular to
$a$-$b$ plane, and compared it with the data in this
work.~\cite{jwenunp} It is shown that there is only anisotropy in
the superconducting state. It will be interesting to see how the
resonance responds to a $c$-axis magnetic field. Another interesting
issue is to search for the Zeeman splitting of the resonance mode
under an external field, which is a good test of whether this is a
singlet-triplet excitation. Zhao {\it et al.}~\cite{zhao-2009} tried
to tackle this problem using a 14.5-T field, but the results are
inconclusive---the resonance in BaFe$_{1.9}$Ni$_{0.1}$As$_2$
broadens in the field, but no clear split was observed, probably due
to the finite resonance width and coarse energy resolution. Qiu {\it
et al.}~\cite{qiu:067008} applied a 7-T magnetic field on
FeTe$_{0.6}$Se$_{0.4}$, but no splitting is visible from their
results; in a more recent experiment, with a larger field (14~T) and
improved background, they were able to resolve the Zeeman splitting,
directly establishing its triplet character~\cite{wbao2}.

In summary, we observed a resonance at $\hbar\Omega_0 \approx 6$~meV
in FeTe$_{0.5}$Se$_{0.5}$ ($T_c=14$~K). The temperature dependence
of the intensity is consistent with the scaling $1-(T/T_c)^{1/2}$. A
7~T magnetic field partially suppresses superconductivity, and
lowers $T_c$ to about 12~K, determined from the bulk susceptibility.
In the field, the resonance starts to appear at the lowered $T_c$,
12~K, with intensity reduced. These results are consistent with the
picture that the resonance is related to quasiparticle scattering in
the superconducting phase, and is reduced when superconductivity
becomes weaker, either by heating or applying an external magnetic
field.

The work at Brookhaven National Laboratory was supported by the U.S.
Department of Energy, Office of Basic Energy Sciences, Division of
Materials Sciences and Engineering, under Contract
No.\,DE-AC02-98CH10886.


\end{document}